\documentclass[12pt,a4paper]{article}
\usepackage{amsmath,amssymb}
\usepackage[nosort]{cite}
\usepackage{graphicx}
\newcommand{\ind}[2]{^{#1}_{\text{#2}}}
\newcommand{\inds}[2]{^{#1}_{\text{\scriptsize{#2}}}}
\newcommand{\indt}[2]{^{#1}_{\text{\tiny{#2}}}}
\newcommand{\ATL}[2]{A^{#1}_{{\rm TL},#2}}
\def\mpi{m_{\pi}}
\def\DP{\Delta\Pi}
\def\Nc{N_{\text c}}
\def\nf{n_{f}}
\def\ds{\displaystyle}

\begin{document}

\begin{center}

{\Large\bf Explicit form of the $R$--ratio of electron--positron annihilation into hadrons

}

\vskip7.5mm

{\large A.V.~Nesterenko}

\vskip5mm

{\small\it Bogoliubov Laboratory of Theoretical Physics,
Joint Institute for Nuclear Research,\\
Dubna, 141980, Russian Federation}

\end{center}

\vskip2.5mm

\noindent
\centerline{\bf Abstract}

\vskip2.5mm

\centerline{\parbox[t]{150mm}{%
The explicit expression for the $R$--ratio of electron--positron
annihilation into hadrons, which properly accounts for all the effects
due to continuation of the spacelike perturbative results into the
timelike domain, is~obtained at an arbitrary loop level. Several
equivalent ways to derive a commonly employed approximation of the
$R$--ratio are recapped and the impact of discarded in the latter
higher--order \mbox{$\pi^2$--terms} on the evaluation of the strong
running coupling is~elucidated. The~obtained results substantially
facilitate the theoretical study of electron--positron annihilation
into hadrons and the related strong interaction processes.
\\[2.5mm]
}}

\vskip10mm

\section{Introduction}
\label{Sect:Intro}

A~broad range of topics in particle physics (including
electron--positron annihilation into hadrons, inclusive hadronic decays
of $\tau$~lepton and $Z$~boson, hadronic contributions to such precise
electroweak observables as the muon anomalous magnetic moment and the
running of the electromagnetic coupling) is inherently based on the
hadronic vacuum polarization function~$\Pi(q^2)$, the Adler
function~$D(Q^2)$, and the function~$R(s)$. The theoretical study of
the aforementioned strong interaction processes factually represents a
decisive consistency test of Quantum Chromodynamics~(QCD) and entire
Standard Model, that puts stringent constraints on a possible new
fundamental physics beyond the latter. Additionally, a majority of the
processes on hand are of a direct pertinence to the numerous ongoing
research programs and the future collider projects, for example, the
first phases of Future Circular Collider~(FCC)~\cite{FCC} and Circular
Electron--Positron Collider~(CEPC)~\cite{CEPC}, International Linear
Collider~(ILC)~\cite{ILC}, Compact Linear Collider~(CLIC)~\cite{CLIC},
as well as E989~experiment at Fermilab~\cite{E989}, E34~experiment
at~\mbox{J--PARC}~\cite{E34}, MUonE~project~\cite{MUonE}, and many
others.

Basically, over the past decades the QCD perturbation theory still
remains to be an essential tool for the theoretical investigation of
the strong interaction processes. However, leaving aside the
irrelevance of perturbative approach to the low--energy hadron
dynamics\footnote{As~an example, one may mention here such
nonperturbative methods, as lattice simulations~\cite{Latt1, Latt2},
Dyson--Schwinger equations~\cite{DSE1, DSE2}, analytic
gauge--invariant~QCD~\cite{Fried}, Nambu--Jona--Lasinio
model~\cite{NJL} and its extended form~\cite{xNJL}, nonlocal chiral
quark model~\cite{NLcQM}, as well as many others.}, it is necessary to
outline that the QCD perturbation theory and the renormalization
group~(RG) method can be directly applied to the study of the strong
interactions in the spacelike (Euclidean) domain only, whereas the
proper description of hadron dynamics in the timelike (Minkowskian)
domain substantially relies on the corresponding dispersion
relations\footnote{It~is worthwhile to note also that the pattern of
applications of dispersion relations in theoretical particle physics is
quite diverse and includes such issues as, for example, the assessment
of hadronic light--by--light scattering~\cite{Colangelo}, the accurate
evaluation of parameters of hadronic resonances~\cite{DRRes}, the
refinement of chiral perturbation theory~\cite{DRChPT, Passemar}, and
many others.}. In~particular, the latter provide the physically
consistent way to interrelate the ``timelike'' experimentally
measurable observables (such~as the \mbox{$R$--ratio} of
electron--positron annihilation into hadrons) with the ``spacelike''
theoretically computable quantities (such~as the Adler function).

In~general, the calculation of the \emph{explicit} expression for the
function~$R(s)$, which thoroughly incorporates all the effects of
continuation of perturbative results from spacelike to timelike domain,
represents quite a demanding task. Eventually, this fact forces one to
either employ the numerical computation methods or resort to an
approximate form of the \mbox{$R$--ratio}, specifically, its truncated
re--expansion at high energies. However, the former requires a lot of
computational resources and becomes rather sophisticated at the higher
loop levels, whereas the latter generates an infinite number of the
so--called \mbox{$\pi^2$--terms}, which may not necessarily be small
enough to be safely neglected at the higher orders and may produce a
considerable effect on~$R(s)$ even at high energies (a~detailed
discussion of these issues can be found in, e.g., Refs.~\cite{Penn,
Rad82, KP82, Bj89, ProsperiAlpha} as well as Refs.~\cite{Book, EPJC77}
and references therein).

The primary objective of the paper is to obtain, at an arbitrary loop
level, the explicit expression for the $R$--ratio of electron--positron
annihilation into hadrons, which properly accounts for all the effects
due to continuation of the spacelike perturbative results into the
timelike domain. It~is also of an apparent interest to elucidate the
impact of the higher--order $\pi^2$--terms, discarded in a commonly
employed approximation of~$R(s)$, on the evaluation of the strong
running coupling.

The layout of the paper is as follows. Section~\ref{Sect:Methods}
delineates the basic dispersion relations for the functions on~hand,
expounds the perturbative expressions for the hadronic vacuum
polarization function and the Adler function, and describes various
ways to handle the function~$R(s)$. In~Sect.~\ref{Sect:Results} the
explicit expression for the $R$--ratio, which entirely embodies all the
effects of continuation of perturbative results from spacelike to
timelike domain, is obtained at an arbitrary loop level, the equivalent
ways to derive an approximate form of the function~$R(s)$ are recapped,
and the impact of ignored in the latter higher--order
\mbox{$\pi^2$--terms} on the evaluation of the strong running coupling
is elucidated. Section~\ref{Sect:Concl} summarizes the basic results.
Appendix~\ref{Sect:CoeffsRG} contains the RG~relations for the hadronic
vacuum polarization function perturbative expansion coefficients.

\section{Methods}
\label{Sect:Methods}

\subsection{Basic dispersion relations}
\label{Sect:DispRels}

To begin, let us briefly expound the essentials of dispersion relations
for the hadronic vacuum polarization function~$\Pi(q^2)$, the Adler
function~$D(Q^2)$, and the function~$R(s)$ (the~detailed description of
this issue can be found in,~e.g., Chap.~1 of Ref.~\cite{Book} and
references therein). As~mentioned earlier, the theoretical study of a
certain class of the strong interaction processes is based on the
hadronic vacuum polarization function~$\Pi(q^2)$, which is defined as
the scalar part of the hadronic vacuum polarization tensor
\begin{equation}
\label{P_Def}
\Pi_{\mu\nu}(q^2) = i\!\int\!d^4x\,e^{i q x} \bigl\langle 0 \bigl|\,
T\!\left\{J_{\mu}(x)\, J_{\nu}(0)\right\} \bigr| 0 \bigr\rangle =
\frac{i}{12\pi^2} (q_{\mu}q_{\nu} - g_{\mu\nu}q^2) \Pi(q^2).
\end{equation}
As discussed in, e.g., Ref.~\cite{Feynman}, for the processes involving
final state hadrons the function~$\Pi(q^2)$~(\ref{P_Def}) possesses the
only cut along the positive semiaxis of real~$q^2$ starting at the
hadronic production threshold~$q^2 \ge 4\mpi^2$, that~implies
\begin{equation}
\label{P_Disp}
\DP(q^2\!,\, q_0^2) = (q^2 - q_0^2) \int\limits_{4\mpi^2}^{\infty}
\frac{R(\sigma)}{(\sigma-q^2)(\sigma-q_0^2)}\, d\sigma,
\end{equation}
where~$\DP(q^2\!,\, q_0^2) = \Pi(q^2) - \Pi(q_0^2)$~and
\begin{equation}
\label{R_Def}
R(s) = \frac{1}{2 \pi i} \lim_{\varepsilon \to 0_{+}}
\Bigl[\Pi(s+i\varepsilon) - \Pi(s-i\varepsilon)\Bigr]\! =
\frac{1}{\pi}\,{\rm Im} \!\lim_{\varepsilon \to 0_{+}}\!
\Pi(s+i\varepsilon).
\end{equation}
The function~$R(s)$~(\ref{R_Def}) is commonly identified with the
$R$--ratio of electron--positron annihilation into hadrons $R(s) =
\sigma(e^{+}e^{-} \!\to \text{hadrons}; s)/\sigma(e^{+}e^{-} \!\to
\mu^{+}\mu^{-}; s)$, where \mbox{$s=q^2>0$} stands for the timelike
kinematic variable, namely, the center--of--mass energy squared.

\begin{figure}[t]
\centerline{\includegraphics[width=77.5mm,clip]{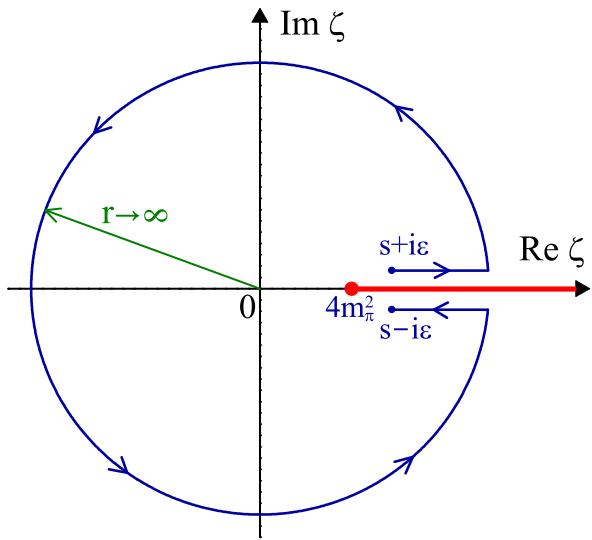}}
\caption{The integration contour in the complex $\zeta$--plane in
Eq.~(\ref{R_Disp2}). The physical cut $\zeta \ge 4\mpi^2$ of the Adler
function $D(-\zeta)$~(\ref{Adler_Def}) is shown along the positive
semiaxis of real~$\zeta$.}
\label{Plot:Contour}
\end{figure}

For~practical purposes it proves to be convenient to deal with the
Adler function~\cite{Adler}
\begin{equation}
\label{Adler_Def}
D(Q^2) = - \frac{d\, \Pi(-Q^2)}{d \ln Q^2},
\end{equation}
where~$Q^2=-q^2>0$ denotes the spacelike kinematic variable. The
corresponding dispersion relation~\cite{Adler}
\begin{equation}
\label{Adler_Disp}
D(Q^2) = Q^2 \int\limits_{4\mpi^2}^{\infty}
\frac{R(\sigma)}{(\sigma+Q^2)^2}\, d\sigma
\end{equation}
directly follows from Eqs.~(\ref{P_Disp}) and~(\ref{Adler_Def}) and
enables one to extract the experimental prediction for the Adler
function from the respective measurements of the~$R$--ratio. In~turn,
the theoretical expression for the function~$R(s)$ can be obtained by
integrating Eq.~(\ref{Adler_Def}) in finite limits, namely~\cite{Rad82,
KP82}
\begin{equation}
\label{R_Disp2}
R(s) =  \frac{1}{2 \pi i} \lim_{\varepsilon \to 0_{+}}
\int\limits_{s + i \varepsilon}^{s - i \varepsilon}
D(-\zeta)\,\frac{d \zeta}{\zeta},
\end{equation}
where the integration contour lies in the region of analyticity of the
integrand, see Fig.~\ref{Plot:Contour}. In~particular,
Eq.~(\ref{R_Disp2}) relates the $R$--ratio to the Adler function,
thereby providing a native way to properly account for all the effects
due to continuation of the spacelike theoretical results into the
timelike domain. At~the same time, Eq.~(\ref{Adler_Def}) additionally
supplies the relation, which expresses the hadronic vacuum polarization
function in terms of the Adler function\footnote{It~is worthwhile to
note here that the calculation of the hadronic vacuum polarization
function~$\Pi(q^2)$ by integration of the Adler function~$D(Q^2)$ was
also described in, e.g., Refs.~\cite{Penn, Georgi73}.},
specifically~\cite{Pivovarov91}
\begin{equation}
\label{P_Disp2}
\DP(-Q^2\!,\, -Q_0^2) = - \int\limits_{Q_0^2}^{Q^2} D(\zeta)
\frac{d \zeta}{\zeta},
\end{equation}
where $Q^{2}>0$ and~$Q_{0}^{2}>0$ denote, respectively, the spacelike
kinematic variable and the subtraction point.

In~fact, Eqs.~(\ref{P_Disp})--(\ref{P_Disp2}) form the complete set of
relations, which mutually express the functions~$\Pi(q^2)$, $R(s)$,
and~$D(Q^2)$ in terms of each other, and their derivation, being based
only on the kinematics of the process on hand, requires neither
additional approximations nor model--dependent phenomenological
assumptions. In~turn, the dispersion
relations~(\ref{P_Disp})--(\ref{P_Disp2}) impose a number of stringent
physical nonperturbative restrictions on the functions~$\Pi(q^2)$,
$R(s)$, and~$D(Q^2)$, that should certainly be accounted for when one
reaches the limits of applicability of perturbative approach. It~is
worth mentioning also that the dispersively improved perturbation
theory~(DPT)~\cite{Book, DPT1, DPT2, JPG42} (its~preliminary
formulation was discussed in Ref.~\cite{DPTPrelim}) conflates the
foregoing nonperturbative constraints with corresponding perturbative
input in a self--consistent way, thereby enabling one to overcome some
inherent difficulties of the perturbative approach to~QCD and extending
its applicability range towards the infrared domain, see, in
particular, Chaps.~4 and~5 of Ref.~\cite{Book} and references therein
for the details.

The nonperturbative aspects of the strong interactions will be
disregarded in what follows and the rest of the paper will be primarily
focused on the theoretical description of the \mbox{$R$--ratio} of
electron--positron annihilation into hadrons at intermediate and high
energies, that, in turn, allows one to safely neglect the effects due
to the masses of the involved particles. At~the same time, one has to
be aware that in the limit of~$\mpi=0$ some of the aforementioned
nonperturbative restrictions on the functions~$\Pi(q^2)$, $R(s)$,
and~$D(Q^2)$, which play a substantial role at low energies, appear to
be~lost (a~discussion of the effects due to the nonvanishing hadronic
production threshold on the infrared behavior of the functions on hand
can be found in Refs.~\cite{Book, DPT1, DPT2, JPG42, DPTPrelim,
QCD8245}).

\clearpage

\subsection{Perturbative expressions for~$\Pi(q^2)$ and~$D(Q^2)$}
\label{Sect:PDpert}

In the framework of perturbation theory the hadronic vacuum
polarization function~(\ref{P_Def}) can be represented~as
\begin{equation}
\label{PpertDef}
\Pi^{(\ell)}\bigl(q^2,\mu^2,a\ind{}{s}\bigr) =
\sum_{j=0}^{\ell}\Bigl[a\ind{(\ell)}{s}(\mu^2)\Bigr]^{j}
\sum_{k=0}^{j+1}\Pi_{j,k}\,\ln^{k}\!\!\left(\frac{\mu^2}{-q^2}\right)\!,
\qquad
q^2 \to -\infty.
\end{equation}
In this equation $\ell$~denotes the loop level, $q^2<0$~stands for the
spacelike kinematic variable, $\mu^2>0$~is the renormalization scale,
$a\ind{(\ell)}{s}(\mu^2) =
\alpha\ind{(\ell)}{s}(\mu^2)\beta_{0}/(4\pi)$ denotes the so--called
QCD~couplant, $\beta_0 = 11 - 2\nf/3$ is the one--loop $\beta$~function
perturbative expansion coefficient, $\nf$~stands for the number of
active flavors, the~common prefactor $\Nc\sum_{f=1}^{\nf} Q_{f}^{2}$ is
omitted throughout, \mbox{$\Nc=3$}~denotes the number of colors, and
$Q_{f}$~stands for the electric charge of $f$--th quark. The hadronic
vacuum polarization function~(\ref{PpertDef}) satisfies the
renormalization group equation
\begin{equation}
\label{RGeqnPgen}
\left[\frac{\partial}{\partial\ln\mu^2} +
\frac{\partial a\ind{}{s}(\mu^2)}{\partial\ln\mu^2}
\frac{\partial}{\partial a\ind{}{s}}\right]\!
\Pi\bigl(q^2,\mu^2,a\ind{}{s}\bigr) = \gamma\bigl(a\ind{}{s}\bigr),
\end{equation}
with~$\gamma\bigl(a\ind{}{s}\bigr)$ being the corresponding anomalous
dimension. At~the $\ell$--loop level the perturbative expression for
the latter takes the following form
\begin{equation}
\label{GpertDef}
\gamma^{(\ell)}\bigl(a\ind{}{s}\bigr) =
\sum_{j=0}^{\ell}\gamma_{j}\Bigl[a\ind{(\ell)}{s}(\mu^2)\Bigr]^{j}.
\end{equation}
Recall that at any given order~$j \ge 1$ the perturbative
coefficients~$\Pi_{j,k}$ $(k=1,\ldots,j+1)$ entering
Eq.~(\ref{PpertDef}) can be expressed in terms of the
coefficients~$\gamma_{i}$ $(i=1,\ldots,j)$ and (if~$j \ge 2$)
$\Pi_{i,0}$ $(i=1,\ldots,j-1)$ by making use of~Eq.~(\ref{RGeqnPgen}),
see, e.g., Ref.~\cite{BCK0912} and references therein. Specifically,
since the renormalization group equation for the $\ell$--loop
perturbative QCD couplant~$a\ind{(\ell)}{s}(\mu^2)$ reads
\begin{equation}
\label{RGeqnAgen}
\frac{\partial a\ind{(\ell)}{s}(\mu^2)}{\partial\ln\mu^2} =
-\sum_{i=0}^{\ell-1}B_{i}\Bigl[a\ind{(\ell)}{s}(\mu^2)\Bigr]^{i+2},
\qquad
B_{i} = \frac{\beta_{i}}{\beta_{0}^{i+1}},
\end{equation}
Eq.~(\ref{RGeqnPgen}) can be represented~as
\begin{align}
\label{RGeqnP}
& \sum_{j=0}^{\ell}\Bigl[a\ind{(\ell)}{s}(\mu^2)\Bigr]^{j}\!
\left\{\!\left[
\sum_{k=0}^{j+1}k\Pi_{j,k}\ln^{k-1}\!\!\left(\frac{\mu^2}{-q^2}\right)\!
\right]\! - \gamma_{j} \right\}
= \nonumber \\[1.5mm]
& = \left\{
\sum_{j=0}^{\ell}j\Bigl[a\ind{(\ell)}{s}(\mu^2)\Bigr]^{j-1}
\sum_{k=0}^{j+1}\Pi_{j,k}\ln^{k}\!\!\left(\frac{\mu^2}{-q^2}\right)\!\!
\right\}
\left\{
\sum_{i=0}^{\ell-1}B_{i}\Bigl[a\ind{(\ell)}{s}(\mu^2)\Bigr]^{i+2}
\right\}\!.
\end{align}
In~particular, at the first few loop levels Eq.~(\ref{RGeqnP}) yields
$\Pi_{0,1} = \gamma_{0}$, $\Pi_{1,1} = \gamma_{1}$, $\Pi_{2,1} =
\Pi_{1,0} + \gamma_{2}$, \mbox{$\Pi_{2,2} = \gamma_{1}/2$}
($\Pi_{j,j+1}=0$ for~$j \ge 1$), see also Ref.~\cite{BCK0912}. At~the
higher loop levels the corresponding relations for the
coefficients~$\Pi_{j,k}$~(\ref{PpertDef}), which will be needed for the
purposes of Sect.~\ref{Sect:Disc}, become rather cumbrous and are
gathered in Appendix~\ref{Sect:CoeffsRG}.

As mentioned earlier, in practice it is convenient to employ the Adler
function~(\ref{Adler_Def}), which is defined as the logarithmic
derivative of the hadronic vacuum polarization function~(\ref{P_Def}).
Specifically, Eq.~(\ref{PpertDef}) implies that at the $\ell$--loop
level the perturbative expression for the Adler
function~(\ref{Adler_Def}) reads
\begin{equation}
\label{DpertMu}
D^{(\ell)}\bigl(Q^2,\mu^2,a\ind{}{s}\bigr) =
\sum_{j=0}^{\ell}\Bigl[a\ind{(\ell)}{s}(\mu^2)\Bigr]^{j}
\sum_{k=0}^{j+1}k\Pi_{j,k}\,\ln^{k-1}\!\!\left(\frac{\mu^2}{Q^2}\right)\!,
\qquad
Q^2 \to \infty.
\end{equation}
Then, the native choice of the renormalization scale~$\mu^2=Q^2$
casts Eq.~(\ref{DpertMu}) to a well--known form~($\Pi_{0,1}=1$)
\begin{equation}
\label{DpertDef}
D^{(\ell)}(Q^2) =
\sum_{j=0}^{\ell}\Pi_{j,1}\Bigl[a\ind{(\ell)}{s}(Q^2)\Bigr]^{j} =
1 + d^{(\ell)}(Q^2),
\end{equation}
where
\begin{equation}
\label{DSCpertDef}
d^{(\ell)}(Q^2) = \sum_{j=1}^{\ell} d_{j}
\Bigl[a\ind{(\ell)}{s}(Q^2)\Bigr]^{j}\!,
\qquad
d_{j} =\Pi_{j,1}
\end{equation}
stands for the $\ell$--loop strong correction. For~example, at the
one--loop level ($\ell=1$) the Adler function~(\ref{DpertDef}) takes
a quite simple form
\begin{equation}
\label{Dpert1L}
D^{(1)}(Q^2) =
1 + d_{1}a\ind{(1)}{s}(Q^2),
\qquad
a\ind{(1)}{s}(Q^2) = \frac{1}{\ln(Q^2/\Lambda^2)},
\end{equation}
where~$d_{1} = 4/\beta_{0}$ and~$\Lambda$ stands for the QCD scale
parameter. As~for the higher loop levels, the solution to the
renormalization group equation for the QCD couplant~(\ref{RGeqnAgen})
can be represented as the double sum
\begin{equation}
\label{AItGen}
a\ind{(\ell)}{s}(Q^2) =
\sum_{n=1}^{\ell}\sum_{m=0}^{n-1} b^{m}_{n}\,
\frac{\ln^{m}(\ln z)}{\ln^n z},
\qquad
z=\frac{Q^2}{\Lambda^2},
\end{equation}
where~$b^{m}_{n}$ (the integer superscript~$m$ is not to be confused
with respective power) denotes the combination of the $\beta$~function
perturbative expansion coefficients (in~particular,
\mbox{$b^{0}_{1}=1$}, $b^{0}_{2}=0$,
$b^{1}_{2}=-\beta_{1}/\beta_{0}^{2}=-B_{1}$, see,~e.g., Appendix~A of
Ref.~\cite{Book}). Hence, the $\ell$--loop strong correction to the
Adler function~(\ref{DSCpertDef}) can also be represented~as
\begin{equation}
\label{DSCtbt}
d^{(\ell)}(Q^2) = \!\!\sum_{j=1}^{\ell}\! d_{j}\!
\sum_{n_{1}=1}^{\ell}\ldots\sum_{n_{j}=1}^{\ell}
\sum_{m_{1}=0}^{n_{1}-1}\ldots\sum_{m_{j}=0}^{n_{j}-1}
\!\left(\prod_{i=1}^{j}b^{m_{i}}_{n_{i}}\right)\!
\frac{\ln^{m_{1}+\ldots+m_{j}}(\ln z)}{\ln^{n_{1}+\ldots+n_{j}}z},
\end{equation}
that appears to be technically more appropriate for the purposes of
Sect.~\ref{Sect:Rexpl}.

\subsection{Various ways to handle~$R(s)$}
\label{Sect:Rpert}

As outlined in Sect.~\ref{Sect:DispRels}, the $R$--ratio of
electron--positron annihilation into hadrons can be calculated by
making use of relation~(\ref{R_Disp2}). In~the massless limit one can
cast the latter~to (see also Ref.~\cite{APT0a})
\begin{equation}
\label{Rprop}
R^{(\ell)}(s) = 1 + r^{(\ell)}(s),
\qquad
r^{(\ell)}(s) =
\int\limits_{s}^{\infty}\!\rho^{(\ell)}(\sigma)\,
\frac{d \sigma}{\sigma}.
\end{equation}
In this equation
\begin{equation}
\label{RhoDef}
\rho^{(\ell)}(\sigma) =
\frac{1}{2 \pi i} \lim_{\varepsilon \to 0_{+}}
\Bigl[d^{(\ell)}(-\sigma - i \varepsilon) -
d^{(\ell)}(-\sigma + i \varepsilon)\Bigr]
\end{equation}
is the corresponding spectral function and~$d^{(\ell)}(Q^2)$ stands for
the $\ell$--loop strong correction to the Adler
function~(\ref{DSCpertDef}). As~noted earlier, only perturbative
contributions\footnote{A~discussion of the intrinsically
nonperturbative terms in Eq.~(\ref{RhoDef}) can be found in, e.g.,
Refs.~\cite{PRD6264, Review, MPLA1516, APTCSB} and~\cite{12dAnQCD,
3dAnQCD}.} are retained in Eq.~(\ref{RhoDef}) herein, that, in turn,
makes Eq.~(\ref{Rprop}) identical to that of both the massless limit of
the aforementioned DPT~\cite{Book, DPT1, DPT2, JPG42} and the analytic
approach~\cite{APT0a, APT0b} (some of its recent applications can be
found in, e.g., Refs.~\cite{APT3, APT4, APT1, APT2, APT5, APT6, APT7,
APT9}). At the one--loop level ($\ell=1$) the perturbative spectral
function~$\rho^{(1)}(\sigma)$ can easily be calculated by making use of
Eqs.~(\ref{RhoDef}) and~(\ref{Dpert1L}), namely
\begin{equation}
\label{RhoPert1L}
\rho^{(1)}(\sigma) = \frac{d_{1}}{y^2+\pi^2},
\qquad
y=\ln\biggl(\!\frac{\sigma}{\Lambda^2}\!\biggr).
\end{equation}
In turn, its integration~(\ref{Rprop}) leads to a well--known
result\footnote{Note that the ``timelike'' effective couplant
$a\indt{(1)}{TL}(s)$~(\ref{RCTL1L}) has first appeared in
Ref.~\cite{Schrempp80} and only afterwards was obtained in
Refs.~\cite{Rad82, Pivovarov91, APT0a}.} for the function~$R(s)$
\begin{equation}
\label{RCTL1L}
R^{(1)}(s) = 1 + d_{1}a\inds{(1)}{TL}(s),
\qquad
a\inds{(1)}{TL}(s)=
\frac{1}{2} -
\frac{1}{\pi}\arctan\biggl(\frac{\ln w}{\pi}\biggr),
\end{equation}
where~$w=s/\Lambda^2$ and it is assumed that~$\arctan(x)$ is a monotone
nondecreasing function of its argument: $-\pi/2 \le \arctan(x) \le
\pi/2$ for $-\infty < x < \infty$. In~Eq.~(\ref{RCTL1L})
$a\inds{(1)}{TL}(s)$ constitutes the one--loop couplant, which properly
incorporates all the effects of continuation of the perturbative
expression~$a\ind{(1)}{s}(Q^2)$~(\ref{Dpert1L}) from spacelike to
timelike domain. Beyond the \mbox{one--loop} level, though the spectral
function~(\ref{RhoDef}) can still be calculated explicitly at few
lowest orders of perturbation theory in a straightforward way (see,~in
particular, Refs.~\cite{Review, APT4}), its integration~(\ref{Rprop})
can, in general, be performed by making use of the numerical methods
only, see, e.g., Refs.~\cite{CPC, anQCDprog}.

At~the higher--loop levels the explicit calculation of the perturbative
spectral function~(\ref{RhoDef}) represents a rather demanding task,
that eventually forces one to either evaluate~$\rho^{(\ell)}(\sigma)$
in Eq.~(\ref{Rprop}) numerically (that, however, requires a lot of
computation resources and becomes quite sophisticated
as~$\ell$~increases, thereby essentially slowing down the overall
computation process), or resort to an approximate expression for
the~$R$--ratio. In~particular, for the latter purpose one can apply the
Taylor expansion to the spectral
function~$\rho^{(\ell)}(\sigma)$~(\ref{RhoDef}) at large values of its
argument, that ultimately casts Eq.~(\ref{Rprop}) to
(see~Refs.~\cite{Book, EPJC77, RpertReexp} and references therein for
the details)
\begin{align}
\label{RpertReexp}
R^{(\ell)}(s) & = 1 +
\sum_{j=1}^{\ell} d_{j} \Bigl[a^{(\ell)}_{{\rm s}}(|s|)\Bigr]^{j} -
\sum_{j=1}^{\ell} d_{j}
\sum_{n=1}^{\infty}\! \frac{(-1)^{n+1}\pi^{2n}}{(2n+1)!}
\sum_{k_{1}=0}^{\ell-1}
\!\ldots\!
\sum_{k_{2n}=0}^{\ell-1}\!\!
\left(\,\prod_{p=1}^{2n}B_{k_{p}}\!\right)\!\!
\times \nonumber \\[2.5mm] & \hspace{-7.5mm} \times
\!\!\left[\,\prod_{t=0}^{2n-1}\!
\Bigl(j+t+k_{1}+k_{2}+\ldots+k_{t}\Bigr)\!\right]\!\!
\Bigl[a^{(\ell)}_{{\rm s}}(|s|)\Bigr]^{j+2n+k_{1}+k_{2}+\ldots+k_{2n}},
\qquad
\frac{\sqrt{s}}{\Lambda} > \exp\biggl(\frac{\pi}{2}\biggr).
\end{align}
The re--expanded $R$--ratio~(\ref{RpertReexp}) constitutes the sum of
naive continuation~($Q^2=|s|$) of the perturbative expression for the
Adler function~$D^{(\ell)}(Q^2)$~(\ref{DpertDef}) into the timelike
domain (the~first two terms on its right--hand side) and an infinite
number of the so--called \mbox{$\pi^2$--terms}. As~demonstrated in
Refs.~\cite{Book, EPJC77}, Eq.~(\ref{RpertReexp}) can provide quite
accurate approximation of the~\mbox{$R$--ratio}~(\ref{Rprop})
for~$\sqrt{s}/\Lambda > \exp(\pi/2) \simeq 4.81$, but only if one
retains sufficiently many expansion terms on its right--hand side.
However, the re--expansion~(\ref{RpertReexp}) is commonly truncated at
a given order~$\ell$, that~yields
\begin{equation}
\label{Rappr}
R^{(\ell)}_{{\rm appr}}(s) = 1 + r\ind{(\ell)}{appr}(s),
\qquad
r^{(\ell)}_{{\rm appr}}(s) =
\sum_{j=1}^{\ell} r_{j} \Bigl[a\ind{(\ell)}{s}(|s|)\Bigr]^{j},
\qquad
r_{j} = d_{j} - \delta_{j},
\end{equation}
where $d_{j}$ denote the Adler function perturbative expansion
coefficients~(\ref{DSCpertDef}) and $\delta_{j}$~incorporate the
contributions of the kept~$\pi^2$--terms. Specifically, at the first
two orders the coefficients $\delta_{j}$~(\ref{Rappr}) vanish
(i.e.,~$\delta_{1} = 0$ and~$\delta_{2} = 0$), whereas at the
higher--loop levels~\cite{Bj89, ProsperiAlpha, KS95, Book, EPJC77}
\begin{equation}
\label{Delta345}
\delta_{3} \!=\! \frac{\pi^2}{3} d_{1},
\quad
\delta_{4} \!=\! \frac{\pi^2}{3}\biggl(\frac{5}{2} d_{1} B_{1} +
3 d_{2}\!\biggr)\!,
\quad
\delta_{5} \!=\!
\frac{\pi^2}{3} \biggl[ \frac{3}{2} d_{1} \Bigl( B_{1}^{2} + 2 B_{2} \Bigr)\!
+ 7 d_{2} B_{1} + 6 d_3 \biggr]\!
- \frac{\pi^4}{5} d_{1},\,
\end{equation}
\begin{equation}
\label{Delta6}
\delta_{6} \!=\! \frac{\pi^2}{3}
\biggl[ \frac{7}{2} d_{1} \Bigl( B_{1} B_{2} + B_{3} \Bigr)
+ 4 d_{2} \Bigl( B_{1}^{2} + 2 B_{2} \Bigr)
+ \frac{27}{2} d_3 B_{1} + 10 d_{4} \biggr]\! -
\frac{\pi^4}{5}\! \left( \frac{77}{12} d_{1} B_{1} + 5 d_{2}\! \right)\!\!,
\end{equation}
\vspace*{-2.5mm}
\begin{align}
\label{Delta7}
\delta_{7} & \!=\!
\frac{\pi^2}{3} \Biggl[
4 d_{1} \!\left(\! B_{1} B_{3} + \frac{1}{2} B_{2}^{2} + B_{4} \!\right) +
9 d_{2} \Bigl( B_{1} B_{2} + B_{3} \Bigr)
% + \nonumber \\[1.5mm] &
+ \frac{15}{2} d_{3} \Bigl( B_{1}^{2} + 2 B_{2} \Bigr) +
\nonumber \\[1.5mm] & +
22 d_{4} B_{1} + 15 d_{5} \Biggr]\!
- \frac{\pi^4}{5} \left[
\frac{5}{6} d_{1} \Bigl( 17 B_{1}^{2} + 12 B_{2} \Bigr)
+ \frac{57}{2} d_{2} B_{1} + 15 d_{3} \right]\! + \frac{\pi^6}{7} d_1,
\end{align}
\vspace*{-5mm}
\begin{align}
\label{Delta8}
\delta_{8} & \!=\!
\frac{\pi^2}{3} \Biggl[
\frac{9}{2} d_{1} \Bigl( B_{1} B_{4} + B_{2} B_{3} + B_{5} \Bigr)
% + \nonumber \\[1.5mm] &
+ 10 d_{2} \!\left(\! B_{1} B_{3} + \frac{1}{2} B_{2}^{2} + B_{4}\! \right)
+ \frac{33}{2} d_{3} \Bigl( B_{1} B_{2} + B_{3} \Bigr)
+ \nonumber \\[1.5mm] &
+ 12 d_{4} \Bigl( B_{1}^{2} + 2 B_{2} \Bigr) +
\frac{65}{2} d_{5} B_{1} + 21 d_{6} \Biggr]\!
% - \nonumber \\[1.5mm] &
- \frac{\pi^4}{5} \Biggl[
\frac{15}{8} d_{1} \Bigl( 7 B_{1}^{3} + 22 B_{1} B_{2} + 8 B_{3} \Bigr)
+ \nonumber \\[1.5mm] &
+ \frac{5}{12} d_{2} \Bigl( 139 B_{1}^{2} + 96 B_{2} \Bigr)
+ \frac{319}{4} d_{3} B_{1} + 35 d_{4} \Biggr]\!
% + \nonumber \\[1.5mm] &
+ \frac{\pi^6}{7} \left( \frac{223}{20} d_{1} B_{1} + 7 d_{2} \right)\!.
\end{align}
The~higher--order coefficients~$\delta_{j}$~(\ref{Rappr}) can be found
in Appendix~C of Ref.~\cite{Book}. It~is worthwhile to recall here that
the calculation of discontinuity~(\ref{R_Def}) of the
expression~(\ref{PpertDef}) with subsequent assignment of the
renormalization scale~$\mu^2=|s|$ also leads to the result identical to
Eq.~(\ref{Rappr}), see Sect.~\ref{Sect:Disc} for a discussion of this
issue.

At~the same time, it is necessary to outline that the approximation of
the \mbox{$R$--ratio} in the form of Eq.~(\ref{Rappr}) has certain
shortcomings, see, e.g., Refs.~\cite{Penn, Rad82, KP82, Bj89,
ProsperiAlpha}. In~particular, the expression~(\ref{Rappr}) discards
all the higher--order $\pi^2$--terms\footnote{As~argued in
Ref.~\cite{Penn}, the representation of the $R$--ratio in the form of
power series in a parameter, which differs
from~$a\ind{(\ell)}{s}(|s|)$, allows one to account for some of the
higher--order $\pi^2$--terms ignored in the
approximation~(\ref{Rappr}), but only partially.}, though the latter,
being not necessarily negligible due to a rather large values of the
coefficients~$\delta_j$, may produce a sizable effect even at high
energies. Moreover, the approximation~$R^{(\ell)}_{{\rm
appr}}(s)$~(\ref{Rappr}) becomes quite inaccurate when~$s$ approaches
the lower bound of its validity range~$\sqrt{s}/\Lambda > \exp(\pi/2)
\simeq 4.81$ and its loop convergence is worse than that of the
expression~(\ref{Rprop}), see Refs.~\cite{Penn, Rad82, KP82, Bj89,
ProsperiAlpha, Book, EPJC77} and references therein for a detailed
discussion of these issues.

It~is worthwhile to mention also that the explicit expression for the
perturbative spectral function~$\rho^{(\ell)}(\sigma)$~(\ref{RhoDef})
has recently been derived at an arbitrary loop level
in~Refs.~\cite{Book, EPJC77}. On~the one hand, the obtained expression
for~$\rho^{(\ell)}(\sigma)$ drastically simplifies the computation of
the $R$--ratio~(\ref{Rprop}), but on the other hand it still requires
one to apply the methods of numerical integration, that may in general
be somewhat effortful.

\section{Results and Discussion}
\label{Sect:Results}

\subsection{Explicit form of the~$R$--ratio}
\label{Sect:Rexpl}

The explicit expression for the~$R$--ratio of electron--positron
annihilation into hadrons, which properly embodies all the effects of
continuation of perturbative results from spacelike to timelike domain,
can be obtained in the following~way. Namely, for this purpose it is
convenient to employ relation~(\ref{P_Disp2}) and Eq.~(\ref{DpertDef})
and then set~$Q^{2}=-s-i0_{+}$ and~$Q_{0}^{2}=-s+i0_{+}$, that makes
the former identical (up~to a constant factor~$2 \pi i$) to the
relation~(\ref{R_Disp2}). Equivalently, one can also employ
relation~(\ref{P_Disp2}) and Eq.~(\ref{DpertDef}), then
set~$Q^{2}=-s-i0_{+}$ and take its imaginary part, that makes the
former identical (up~to a constant factor~$\pi$) to the
relation~(\ref{R_Def}).

Specifically, at the one--loop level~($\ell=1$) it is straightforward
to demonstrate that relation~(\ref{P_Disp2}) and Eq.~(\ref{Dpert1L})
lead to (see,~e.g., Refs.~\cite{Penn, Pivovarov91, Book, JPG42})
\begin{equation}
\label{DP1L}
\DP^{(1)}(-Q^{2},-Q_{0}^{2}) = -\ln\!\biggl(\frac{Q^2}{Q_0^2}\biggr)
- d_{1}\ln\!\Biggl[\frac{a\ind{(1)}{s}(Q_{0}^{2})}{a\ind{(1)}{s}(Q^{2})}\Biggr]\!,
\end{equation}
where~$d_{1}=4/\beta_{0}$
and~$a\ind{(1)}{s}(Q^{2})=1/\ln(Q^2/\Lambda^2)$. Then, the explicit
expression for the function~$R(s)$ can be obtained from
Eq.~(\ref{DP1L}) by making use of relation~(\ref{R_Def}), namely
\begin{equation}
% \label{R1L}
R^{(1)}(s) = \frac{1}{\pi}\,{\rm Im} \!\lim_{\varepsilon \to 0_{+}}\!
\DP^{(1)}(s+i\varepsilon,-Q_{0}^{2}) =
1 + d_{1}\biggl[\frac{1}{2}-\frac{1}{\pi}\arctan\biggl(\frac{\ln w}{\pi}\biggr)\!\biggr],
\qquad
w=\frac{s}{\Lambda^2},
\end{equation}
that obviously coincides with the result~(\ref{RCTL1L}) obtained
earlier from relation~(\ref{R_Disp2}) and Eq.~(\ref{Dpert1L}).

As~for the higher--loop levels, Eq.~(\ref{DpertDef}) implies that the
right--hand side of relation~(\ref{P_Disp2}) is composed of the terms
of a~form
\begin{equation}
\label{PGT}
-\int\limits_{Q_{0}^{2}}^{Q^{2}}
\frac{\ln^{m}\bigl[\ln(\zeta/\Lambda^2)\bigr]}{\ln^{n}(\zeta/\Lambda^2)}
\frac{d\zeta}{\zeta} =
-\!\!\!\int\limits_{\ln(\ln z_{0})}^{\ln(\ln z)}\!\!\!
e^{-x(n-1)}x^{m} d x,
\end{equation}
where~$z=Q^{2}/\Lambda^{2}$ and~$z_{0}=Q_{0}^{2}/\Lambda^{2}$. Then,
since (up to an insufficient integration constant)
\begin{equation}
\label{Int1}
-\!\!\int\!\!e^{-x(n-1)}x^{m} d x =
\begin{cases}
\ds -\frac{x^{m+1}}{m+1},& \text{if $\, n=1$},\\[4mm]
\ds \frac{\Gamma\bigl[m+1,\,x(n-1)\bigr]}{(n-1)^{m+1}},
\quad& \text{if $\, n \ge 2$}
\end{cases}
\end{equation}
and
\begin{equation}
\label{Int2}
\Gamma(m,x) = (m-1)!\,e^{-x}e_{m-1}(x) = (m-1)!\,e^{-x}
\sum_{k=0}^{m-1}\frac{x^k}{k!},
\end{equation}
one can cast Eq.~(\ref{PGT})~to
\begin{equation}
-\int\limits_{Q_{0}^{2}}^{Q^{2}}
\frac{\ln^{m}\bigl[\ln(\zeta/\Lambda^2)\bigr]}{\ln^{n}(\zeta/\Lambda^2)}
\frac{d\zeta}{\zeta} = J(Q^{2},n,m) - J(Q_{0}^{2},n,m),
\end{equation}
where
\begin{equation}
\label{JDef}
J(Q^{2},n,m) =
\begin{cases}
\ds -\frac{\ln^{m+1}(\ln z)}{m+1},& \text{if $\, n=1$},\\[4mm]
\ds \sum_{k=0}^{m}\frac{m!}{k!}(n-1)^{k-m-1}\,\frac{\ln^{k}(\ln z)}{\ln^{n-1}z},
\quad& \text{if $\, n \ge 2$}.
\end{cases}
\end{equation}
In~Eq.~(\ref{Int1}) $\Gamma(m,x)$ stands for the complementary
(or~``upper'') incomplete gamma function, whereas in Eq.~(\ref{Int2})
$e_{m}(x)$ denotes the exponential sum function, see Ref.~\cite{BE}.
Thus, at the $\ell$--loop level relation~(\ref{P_Disp2}) and
Eq.~(\ref{DpertDef}) yield
\begin{equation}
\label{DPHL}
\DP^{(\ell)}(-Q^{2},-Q_{0}^{2}) = -\ln\!\left(\frac{Q^{2}}{Q_{0}^{2}}\right)
+ \Delta p^{(\ell)}(Q^{2},Q_{0}^{2}),
\end{equation}
where
\begin{equation}
\Delta p^{(\ell)}(Q^{2},Q_{0}^{2}) = \sum_{j=1}^{\ell}d_{j}
\Bigl[
p^{(\ell)}_{j}(Q^{2}) - p^{(\ell)}_{j}(Q_{0}^{2})\Bigr],
\end{equation}
\begin{equation}
\label{PscJL}
p^{(\ell)}_{j}(Q^{2}) =
\sum_{n_{1}=1}^{\ell}\ldots\sum_{n_{j}=1}^{\ell}
\sum_{m_{1}=0}^{n_{1}-1}\ldots\sum_{m_{j}=0}^{n_{j}-1}
\Biggl(\prod_{i=1}^{j}b^{m_{i}}_{n_{i}}\Biggr)
J\Biggl(\!Q^{2},\sum_{i=1}^{j}n_{i},\sum_{i=1}^{j}m_{i}\!\Biggr),
\end{equation}
the coefficients~$d_{j}$ and~$b^{m}_{n}$ are specified in
Eqs.~(\ref{DSCpertDef}) and~(\ref{AItGen}), respectively, and the
function~$J(Q^{2},n,m)$ is defined in Eq.~(\ref{JDef}).

In~turn, the obtained expression~(\ref{DPHL}) implies that at the
higher--loop levels the right--hand side of relation~(\ref{R_Def}) is
comprised of the terms of a~form
\begin{equation}
\label{VnmDef}
\frac{1}{2 \pi i} \lim_{\varepsilon \to 0_{+}}
\Bigl[ \bar a_{n}^{m}(-s-i\varepsilon) - \bar a_{n}^{m}(-s+i\varepsilon)\Bigr]\! =
\frac{1}{\pi}\,{\rm Im} \!\lim_{\varepsilon \to 0_{+}}\!
\bar a_{n}^{m}(-s-i\varepsilon) = V_{n}^{m}(s),
\end{equation}
where
\begin{equation}
\label{RCLnm}
\bar a_{n}^{m}(Q^2) = \frac{\ln^{m}(\ln z)}{\ln^n z},
\qquad
z = \frac{Q^2}{\Lambda^2}.
\end{equation}
The function~$V_{n}^{m}(s)$ appearing in Eq.~(\ref{VnmDef})~reads
\begin{equation}
\label{VnmBar}
V_{n}^{m}(s) =
\begin{cases}
0,\quad& \text{if $\, n=0 \,$ and $\, m = 0$}, \\[1.25mm]
v_{0}^{m}(s),\quad& \text{if $\, n=0 \,$ and $\, m \ge 1$}, \\[1.25mm]
v_{n}^{0}(s),& \text{if $\, n \ge 1 \,$ and $\, m=0$}, \\[1.25mm]
v_{n}^{0}(s) u_{0}^{m}(s) +
u_{n}^{0}(s) v_{0}^{m}(s),\quad& \text{if $\, n \ge 1 \,$ and $\, m \ge 1$},
\end{cases}
\end{equation}
where
\begin{align}
\label{V0mDef}
v_{0}^{m}(s) & = \!\sum\limits_{k=0}^{K(m)}\!\binom{m}{2k+1} (-1)^{k+1}\pi^{2k}
\Bigl[L_{1}(y)\Bigr]^{m-2k-1}\, \Bigl[L_{2}(y)\Bigr]^{2k+1},
\\[1.5mm]
\label{Vn0Def}
v_{n}^{0}(s) & = \frac{1}{(y^2 + \pi^2)^{n}}
\sum_{k=0}^{K(n)} \!\binom{n}{2k+1} (-1)^{k} \pi^{2k} y^{n-2k-1},
\end{align}
\begin{align}
\label{U0mDef}
u_{0}^{m}(s) & = \!\sum\limits_{k=0}^{K(m+1)}\!\binom{m}{2k} (-1)^{k}\pi^{2k}
\Bigl[L_{1}(y)\Bigr]^{m-2k}\, \Bigl[L_{2}(y)\Bigr]^{2k},
\\[1.5mm]
\label{Un0Def}
u_{n}^{0}(s) & = \frac{1}{(y^2 + \pi^2)^{n}}
\sum_{k=0}^{K(n+1)} \!\binom{n}{2k} (-1)^{k} \pi^{2k} y^{n-2k},
\end{align}
\begin{equation}
\label{L12Def}
L_{1}(y) = \ln\!\sqrt{y^{2}+\pi^{2}},
\qquad
L_{2}(y) = \frac{1}{2} - \frac{1}{\pi}\arctan\biggl(\frac{y}{\pi}\biggr),
\end{equation}
\begin{equation}
\binom{n}{m} = \frac{n!}{m!\,(n-m)!},
\qquad
\label{KImDef}
K(n) = \frac{n-2}{2} + \frac{n \;\mbox{mod}\; 2}{2},
\end{equation}
$(n \;\mbox{mod}\; m)$ denotes the remainder on division of~$n$ by~$m$,
and~$y = \ln w = \ln(s/\Lambda^2)$.
The~function~$V_{n}^{m}(s)$~(\ref{VnmBar}) constitutes the
generalization of the function~$v_{n}^{m}(s)$ specified in
Refs.~\cite{Book, EPJC77} and the details of its derivation are quite
similar to those given therein. It~is worthwhile to note that in
Eqs.~(\ref{VnmDef}), (\ref{VnmBar})--(\ref{Un0Def}), and on the
left--hand side of Eq.~(\ref{RCLnm}) the integer superscript~$m$ is not
to be confused with respective power. Thus, at the $\ell$--loop level
relation~(\ref{R_Def}) and Eq.~(\ref{DPHL}) lead to the following
expression for the $R$--ratio of electron--positron annihilation into
hadrons:
\begin{equation}
\label{Rprop2}
R^{(\ell)}(s) = 1 + r^{(\ell)}(s), \qquad
r^{(\ell)}(s) = \sum\limits_{j=1}^{\ell} d_{j}\,\ATL{(\ell)}{j}(s),
\end{equation}
where~$d_{j}$ stand for the Adler function perturbative expansion
coefficients~(\ref{DSCpertDef}),
\begin{equation}
\label{ATLdef}
\ATL{(\ell)}{j}(s) =
\sum_{n_{1}=1}^{\ell}\ldots\sum_{n_{j}=1}^{\ell}
\sum_{m_{1}=0}^{n_{1}-1}\ldots\sum_{m_{j}=0}^{n_{j}-1}
\Biggl(\prod_{i=1}^{j}b^{m_{i}}_{n_{i}}\Biggr)\mbox{\tiny$\,$}
T\mbox{\tiny$\!$}\Biggl(\!s,\sum_{i=1}^{j}n_{i},\sum_{i=1}^{j}m_{i}\!\Biggr)
\end{equation}
denotes the $\ell$--loop $j$--th~order ``timelike'' effective expansion
function (that constitutes the continuation of the $j$--th power of
$\ell$--loop QCD couplant~$\bigl[a\ind{(\ell)}{s}(Q^2)\bigr]^{j}$ into
the timelike domain), the coefficients~$b^{m}_{n}$ are specified in
Eq.~(\ref{AItGen}),
\begin{equation}
\label{TDef}
T(s,n,m) =
\begin{cases}
\ds -V_{0}^{1}(s),& \text{if $\, n=1 \,$ and $\, m=0$},\\[2.5mm]
\ds \sum_{k=0}^{m}\frac{m!}{k!}(n-1)^{k-m-1}\,V_{n-1}^{k}(s),
\quad& \text{if $\, n \ge 2$},
\end{cases}
\end{equation}
and the function~$V_{n}^{m}(s)$ is defined in Eq.~(\ref{VnmBar}).
The~obtained explicit expression for the
\mbox{$R$--ratio}~(\ref{Rprop2})--(\ref{TDef}) properly accounts for
all the effects due to continuation of the spacelike perturbative
results into the timelike domain and, being valid at an arbitrary loop
level, can easily be employed in practical applications.

It is worthwhile to note also that Eq.~(\ref{ATLdef}) certainly
coincides with all five explicit expressions for the
function~$\ATL{(\ell)}{j}(s)$ obtained thus far by making use of the
method described in Sect.~\ref{Sect:Rpert}. Specifically, as mentioned
earlier, the one--loop first--order ($\ell=1$, $j=1$) expansion
function~(\ref{ATLdef}) is identical to Eq.~(\ref{RCTL1L}), which was
obtained for the first time in Ref.~\cite{Schrempp80}, namely
\begin{equation}
\ATL{(1)}{1}(s) = L_{2}(y) = a\inds{(1)}{TL}(s) =
\frac{1}{2} -
\frac{1}{\pi}\arctan\biggl(\frac{y}{\pi}\biggr),
\end{equation}
where~$L_{2}(y)$ is defined in Eq.~(\ref{L12Def}) and~$y = \ln w =
\ln(s/\Lambda^2)$. In~turn, the two--loop first--order
(\mbox{$\ell=2$}, \mbox{$j=1$}) expansion function~(\ref{ATLdef})
\begin{equation}
\label{ATL21expl}
\ATL{(2)}{1}(s) = L_{2}(y) - \frac{B_{1}}{y^2 + \pi^2}
\biggl[1 + L_{1}(y) - yL_{2}(y)\biggr]
\end{equation}
and the two--loop second--order (\mbox{$\ell=2$}, \mbox{$j=2$})
expansion function~(\ref{ATLdef})
\begin{align}
\label{ATL22expl}
\ATL{(2)}{2}(s) & = \frac{1}{y^2 + \pi^2} + \frac{B_{1}}{(y^2 + \pi^2)^2}
\biggl\{L_{2}(y)(y^2-\pi^2) - y\Bigl[1+2L_{1}(y)\Bigr]\!\biggr\}
+ \nonumber \\[2.5mm] &
+ \frac{2}{3} \frac{B_{1}^{2}}{(y^2 + \pi^2)^3}
\Biggl\{\!(3y^2-\pi^2)\biggl[\frac{1}{9} + \frac{1}{3}L_{1}(y) +
\frac{1}{2}\Bigl(L_{1}^{2}(y) - \pi^{2} L_{2}^{2}(y)\!\Bigr)\!\biggr]
- \nonumber \\[2.5mm] &
- y\Bigl(y^2-3\pi^2\Bigr)L_{2}(y)\biggl[\frac{1}{3} + L_{1}(y)\biggr]\!
\Biggr\}
\end{align}
coincide with the corresponding expressions obtained for the first time
in Ref.~\cite{Rad82} and Ref.~\cite{Book}, respectively.
In~Eqs.~(\ref{ATL21expl}) and~(\ref{ATL22expl}) $B_{1} =
\beta_{1}/\beta_{0}^{2}$ and~$L_{1}(y)$ is specified in
Eq.~(\ref{L12Def}). It~is also straightforward to verify that the
three--loop first--order (\mbox{$\ell=3$}, \mbox{$j=1$}) and the
four--loop first--order (\mbox{$\ell=4$}, \mbox{$j=1$}) expansion
functions~(\ref{ATLdef}) are identical to the corresponding expressions
obtained for the first time in~Ref.~\cite{EPJC77}. The~explicit form of
the other functions~$\ATL{(\ell)}{j}(s)$ had remained hitherto
unavailable. In~particular, as noted above, the computation of the
functions~$\ATL{(\ell)}{j}(s)$ and the study of the \mbox{$R$--ratio}
at the first five loop levels\footnote{For the five--loop
$\beta$~function perturbative expansion
coefficient~$\beta_{4}$~(\ref{RGeqnAgen}) its recent
calculation~\cite{Beta5L} is used, whereas for the unavailable yet
five--loop Adler function perturbative expansion
coefficient~$d_{5}$~(\ref{DSCpertDef}) its numerical
estimation~\cite{KS95} is employed.}, which was performed in
Ref.~\cite{EPJC77}, employ the methods of numerical integration.

\subsection{Discussion}
\label{Sect:Disc}

In~fact, the foregoing approximate expression for
the~\mbox{$R$--ratio}~(\ref{Rappr}) can be obtained in several
equivalent ways. Specifically, to derive~$R_{{\rm
appr}}(s)$~(\ref{Rappr}) one can apply the Taylor expansion to the
proper expression for the function~$R(s)$ at high energies, rearrange
the expansion terms in the way described in, e.g.,~Sect.~6.2 of
Ref.~\cite{Book}, and then truncate it at a given order. Alternatively,
one can apply the Taylor expansion to the corresponding spectral
function~$\rho(\sigma)$~(\ref{RhoDef}) at large values of its argument,
perform term--by--term integration~(\ref{Rprop}) of the result [that
yields Eq.~(\ref{RpertReexp})], and then truncate it at a given order.
Additionally, as mentioned earlier, the calculation of
discontinuity~(\ref{R_Def}) of the
expression~$\Pi(q^2,\mu^2,a\ind{}{s})$~(\ref{PpertDef}) with subsequent
assignment of the renormalization scale~\mbox{$\mu^2=|s|$} (that
amounts to an incomplete RG~summation in the timelike domain, see,
e.g., Refs.~\cite{Penn, Pivovarov91} and references therein for a
discussion of this issue) also leads to the result identical to
Eq.~(\ref{Rappr}). In~particular, relation~(\ref{R_Def}) and
Eq.~(\ref{PpertDef}) imply that the coefficients~$\delta_{j}$
entering~$R_{{\rm appr}}(s)$~(\ref{Rappr})~read
\begin{equation}
\label{DeltaGen}
\delta_{1} = 0,
\qquad
\delta_{2} = 0,
\qquad
\delta_{j} = \sum_{k=1}^{K(j)} (-1)^{k+1} \pi^{2k} \Pi_{j,2k+1},
\qquad
j \ge 3,
\end{equation}
with~$K(j)$ being defined in Eq.~(\ref{KImDef}). The~identity of the
coefficients~$\delta_{j}$~(\ref{DeltaGen}) to the corresponding
expressions~(\ref{Delta345})--(\ref{Delta8}) can be demonstrated by
making use of the results presented in Appendix~\ref{Sect:CoeffsRG}.
At~the same time, as noted above, one has to be aware that the
approximation~$R_{{\rm appr}}(s)$~(\ref{Rappr}) becomes quite rough
when the energy scale~$s$ approaches the lower bound of its validity
range~$\sqrt{s}/\Lambda > \exp(\pi/2) \simeq 4.81$ and its loop
convergence is worse than that of the proper expression for
the~\mbox{$R$--ratio}~(\ref{Rprop2}). Moreover, the
\mbox{$\pi^{2}$--terms} ignored in the truncated re--expanded
approximation~$R_{{\rm appr}}(s)$~(\ref{Rappr}) may produce a
considerable effect even at high energies due to a rapid growth of the
higher--order coefficients~$\delta_{j}$, see, e.g., Refs.~\cite{Penn,
Rad82, KP82, Bj89, ProsperiAlpha, Book, EPJC77} and references therein.

\begin{figure}[t]
\centerline{\includegraphics[width=77.5mm,clip]{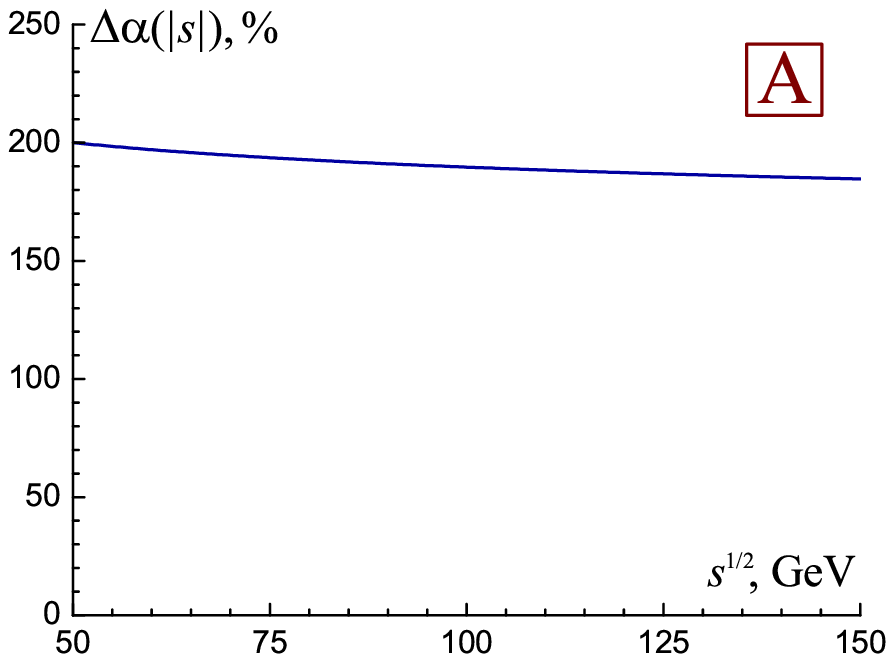}%
\hfill%
\includegraphics[width=77.5mm,clip]{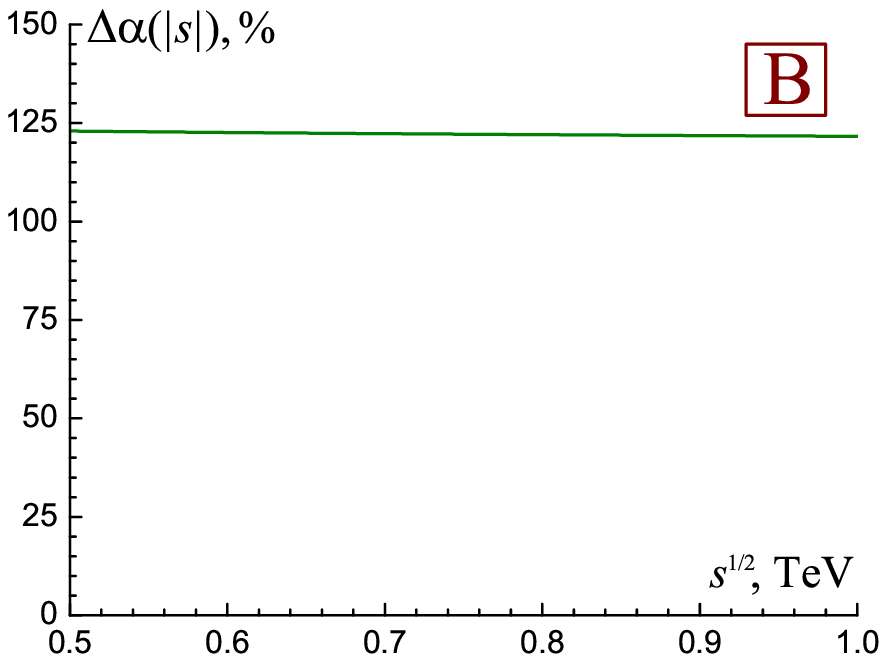}}
\caption{The relative difference~(\ref{DeltaAlpha}) between the effect
of inclusion of the $\pi^{2}$--terms discarded in the four--loop
approximate expression~$R^{(4)}_{{\rm appr}}(s)$~(\ref{Rappr}) and the
effect of inclusion of the five--loop perturbative correction into
Eq.~(\ref{Rappr}) on the resulting value of the strong running coupling
in $\nf=5$~energy range~(plot~A) and the future ILC~experiment energy
range~\cite{ILC}~($\nf=6$, plot~B).}
\label{Plot:AlphaDiff}
\end{figure}

Specifically, to illustrate the convergence of the approximate form of
the $R$--ratio~(\ref{Rappr}), it is worth noting that\footnote{For~the
scheme--dependent perturbative
coefficients~$\beta_{j}$~(\ref{RGeqnAgen})
and~$d_{j}$~(\ref{DSCpertDef}) the $\overline{\rm{MS}}$--scheme is
assumed.} at the four--loop level~($\ell=4$) at the scale of the
$Z$~boson mass ($M_{{\rm Z}}=91.1876\,$GeV~\cite{PDG18}) its
third--order~\mbox{($j=3$)} and fourth--order~\mbox{($j=4$)} terms
comprise, respectively, $34.2\,\%$ and~$8.1\,\%$ of its
second--order~\mbox{($j=2$)}~term. As~for the proper expression for the
$R$--ratio~(\ref{Rprop2}), at the same loop level and energy scale its
third--order~\mbox{($j=3$)} and fourth--order~\mbox{($j=4$)} terms
comprise, respectively, only $1.8\,\%$ and~$0.8\,\%$ of its
second--order~\mbox{($j=2$)}~term, thereby displaying much better
convergence than that of Eq.~(\ref{Rappr}). In~particular, this
exemplifies the fact that (as~argued in~Refs.~\cite{Book, EPJC77}) the
$j$--th order contribution to the function~$R(s)$ appears to be
redistributed over the higher--order terms in its
\mbox{re--expansion}~(\ref{RpertReexp}). The~reported findings also
imply that the uncertainty of the resulting value of the strong running
coupling associated with truncation of the proper expression for the
$R$--ratio~(\ref{Rprop2}) at a given loop level is considerably less
than that of its approximate form~(\ref{Rappr}).

Additionally, Fig.~\ref{Plot:AlphaDiff} presents the relative
difference between the impact of the higher--order $\pi^{2}$--terms
omitted in the four--loop approximation~$R^{(4)}_{{\rm
appr}}(s)$~(\ref{Rappr}) and the impact of the five--loop perturbative
correction to Eq.~(\ref{Rappr}) on the evaluation of the strong running
coupling. Namely, this figure displays the quantity
\begin{equation}
\label{DeltaAlpha}
\Delta\alpha(|s|) = \left|\frac{\bar\alpha\ind{(4)}{s}(|s|) - \alpha\ind{(4)}{s}(|s|)}
{\bar\alpha\ind{(4)}{s}(|s|) - \bar\alpha\ind{(5)}{s}(|s|)}\right|\!\times 100\%,
\end{equation}
where~$\alpha\ind{(\ell)}{s}(|s|)$ and~$\bar\alpha\ind{(\ell)}{s}(|s|)$
stand for the \mbox{$\ell$--loop} strong running coupling evaluated at
the energy scale~$|s|$ by making use of, respectively, the proper
expression~$R^{(\ell)}(s)$~(\ref{Rprop2}) and its approximate
form~$R^{(\ell)}_{{\rm appr}}(s)$~(\ref{Rappr}). In~particular, as one
can infer from Fig.~\ref{Plot:AlphaDiff}, the effect of inclusion of
the \mbox{$\pi^2$--terms} ignored in the four--loop approximate
expression~$R^{(4)}_{{\rm appr}}(s)$~(\ref{Rappr}) on the resulting
value of the strong running coupling (likewise a similar impact on the
\mbox{$R$--ratio} itself, see Refs.~\cite{Book, EPJC77}) is either
prevailing over or comparable to the effect of inclusion of the
five--loop perturbative correction into Eq.~(\ref{Rappr}) even at high
energies. Specifically, the former effect exceeds the latter one by a
factor of two in $\nf=5$~energy range~(plot~A) and by a factor of~$1.2$
in the energy range planned for the future ILC~experiment~\cite{ILC}
($\nf=6$, plot~B).

\begin{table*}[t]
\caption{The values of the strong running coupling and the QCD~scale
parameter at the first five loop levels~$(\ell=1,\ldots,5)$ extracted
from the mean value of the experimental
data~\mbox{$R(s_{0})=1.18$}~\cite{BES02} by making use of the proper
expression for the
\mbox{$R$--ratio}~(\ref{Rprop2})~$\bigl[\alpha\ind{(\ell)}{s}(|s_{0}|),
\, \Lambda^{\!(\ell)}\bigr]$ and its approximate
form~(\ref{Rappr})~$\bigl[\bar\alpha\ind{(\ell)}{s}(|s_{0}|), \,
\bar\Lambda^{\!(\ell)}\bigr]$ at the energy scale
of~$\sqrt{s_{0}}=2\,$GeV.}
\vskip2.5mm
\label{Tab:AlphaDiff}
\begin{tabular*}{\textwidth}{@{\extracolsep{\fill}}cccccc@{\extracolsep{\fill}}}
\hline\\[-4mm]
  & $\ell=1$ & $\ell=2$ & $\ell=3$ & $\ell=4$ & $\ell=5$ \\[2mm]
\hline\\[-4mm]
$\alpha\ind{(\ell)}{s}(|s_{0}|)$     & 0.3283 & 0.3168 & 0.2955 & 0.2955 & 0.2924 \\[2mm]
$\Lambda^{\!(\ell)}$(MeV)     & 238 & 417 & 336 & 331 & 331 \\[2mm]
\hline\\[-4mm]
$\bar\alpha\ind{(\ell)}{s}(|s_{0}|)$ & 0.2827 & 0.2501 & 0.2655 & 0.2881 & 0.3278 \\[2mm]
$\bar\Lambda^{\!(\ell)}$(MeV) & 169 & 263 & 269 & 315 & 408 \\[2mm]
\hline\\[-4mm]
\end{tabular*}
\end{table*}

To elucidate the impact of the higher--order $\pi^2$--terms, discarded
in the approximate expression for the \mbox{$R$--ratio}~(\ref{Rappr}),
on the evaluation of the strong running coupling itself, it is
worthwhile to note the following. At~the energy scale
of~$\sqrt{s_{0}}=2\,$GeV the four--loop strong running coupling assumes
the value~$\alpha\ind{(4)}{s}(|s_{0}|) = 0.2960 \pm 0.0080$, the world
average of the QCD~scale parameter~$\Lambda^{\!(4)} = (332 \pm
17)\,$MeV~\cite{PDG18} being employed. At~the same time, the values of
the strong running coupling and the QCD~scale parameter at the first
five loop levels~$(\ell=1,\ldots,5)$ extracted from the corresponding
mean value of the experimental data~$R(s_{0})=1.18$~\cite{BES02} are
presented in Tab.~\ref{Tab:AlphaDiff}. As~earlier, the
quantities~$\bigl[\alpha\ind{(\ell)}{s}(|s_{0}|), \,
\Lambda^{\!(\ell)}\bigr]$
and~$\bigl[\bar\alpha\ind{(\ell)}{s}(|s_{0}|), \,
\bar\Lambda^{\!(\ell)}\bigr]$ are evaluated by making use of the proper
expression for the \mbox{$R$--ratio}~(\ref{Rprop2}) and its approximate
form~(\ref{Rappr}), respectively. As~one can infer from
Tab.~\ref{Tab:AlphaDiff}, starting from the three--loop level~($\ell
\ge 3$) the inclusion of the higher--order perturbative corrections
into the proper expression for the~$R$--ratio~(\ref{Rprop2}) yields a
rather mild variation of the resulting values of the strong running
coupling and the QCD~scale parameter, thereby reflecting the
aforementioned enhanced convergence of~$R(s)$~(\ref{Rprop2}). On~the
contrary, the use of an approximate form of
the~$R$--ratio~(\ref{Rappr}) results in the values
of~$\bar\alpha\ind{(\ell)}{s}(|s_{0}|)$ and~$\bar\Lambda^{\!(\ell)}$,
which show no sign of convergence and swerve away from the
corresponding values of~$\alpha\ind{(\ell)}{s}(|s_{0}|)$
and~$\Lambda^{\!(\ell)}$. In~turn, this clearly demonstrates the fact
that the approximation~(\ref{Rappr}) is rather rough at the energy
scale on hand and the higher--order $\pi^2$--terms omitted
in~$R^{(\ell)}_{{\rm appr}}(s)$~(\ref{Rappr}) play a significant role
in the evaluation of the strong running coupling and the QCD~scale
parameter.

\section{Conclusions}
\label{Sect:Concl}

The explicit expression for the $R$--ratio of electron--positron
annihilation into hadrons, which properly accounts for all the effects
due to continuation of the spacelike perturbative results into the
timelike domain, is~obtained at an arbitrary loop level
[Eqs.~(\ref{Rprop2})--(\ref{TDef})]. Several equivalent ways to derive
a commonly employed approximation of the $R$--ratio are recapped and
the impact of the discarded in the latter higher--order
\mbox{$\pi^2$--terms} on the evaluation of the strong running coupling
is elucidated. The~obtained results substantially facilitate the
theoretical study of electron--positron annihilation into hadrons and
the related strong interaction processes.

\vskip7.5mm

\noindent
{\large\bf Acknowledgements}

\vskip2.5mm

The author is grateful to Prof.~A.B.~Arbuzov for the stimulating
discussions and useful comments.

\appendix

\section{RG relations for the coefficients~$\Pi_{j,k}$}
\label{Sect:CoeffsRG}

As noted in Sect.~\ref{Sect:PDpert}, at any given order~$j$ the
hadronic vacuum polarization function perturbative expansion
coefficients~$\Pi_{j,k}$ $(k=1,\ldots,j+1)$ entering
Eq.~(\ref{PpertDef}) can be expressed in terms of the
coefficients~$\gamma_{i}$ $(i=0,\ldots,j)$ appearing in
Eq.~(\ref{GpertDef}) and (if~$j \ge 2$)~$\Pi_{i,0}$ $(i=1,\ldots,j-1)$
by making use of the renormalization group equation~(\ref{RGeqnPgen}).
The~corresponding relations for the coefficients~$\Pi_{j,k}$ at the
first eight loop levels $(j=0,\ldots,8)$, which are needed for the
purposes of Sect.~\ref{Sect:Disc}, are presented in the following.

\bigskip

\noindent
First of all, for~$j=0$
\begin{equation}
\Pi_{0,1} = \gamma_{0}.
%\nonumber
\end{equation}
Then, for~$j \ge 1$
\begin{equation}
\Pi_{j,j} = \frac{1}{j}\gamma_{1},
\qquad
\Pi_{j,j+1}=0.
% \nonumber
\end{equation}
In turn, for~$j \ge 2$
\begin{equation}
\Pi_{j,1} = \gamma_{j} + \sum_{k=1}^{j-1}k\,\Pi_{k,0}\,B_{j-k-1},
\qquad
B_{j} = \frac{\beta_{j}}{\beta_{0}^{j+1}}.
% \nonumber
\end{equation}
For~$j \ge 3$
\begin{equation}
% \nonumber
\Pi_{j,2} = \Gamma_{j-1} + \!\sum_{k=1}^{j-2}k(j+k)\Pi_{k,0}\mathfrak{B}_{j-k-2},
\end{equation}
where
\begin{equation}
% \nonumber
\Gamma_{j} = \frac{1}{2}\sum_{k=1}^{j}k\gamma_{k}\,B_{j-k},
\qquad
\mathfrak{B}_{j} = \frac{1}{4} \sum_{k=0}^{j}B_{k}B_{j-k}.
\end{equation}
For~$j \ge 4$
\begin{equation}
% \nonumber
\Pi_{j,3} = \frac{1}{3}\sum_{k=1}^{j-2}k(j+k)\Pi_{k,1}\mathfrak{B}_{j-k-2}.
\end{equation}
The rest of the relations are given below:
\begin{equation}
% \nonumber
\Pi_{5,4} =
\Pi_{1,0}
+ \frac{13}{12}\gamma_{1}B_{1}
+ \gamma_{2}.
\end{equation}

\begin{align}
\Pi_{6,4} & =
\frac{77}{12}\Pi_{1,0}B_{1}
+ 5\Pi_{2,0}
+ \gamma_{1}\biggl(\frac{35}{24}B_{1}^{2} + \frac{3}{2}B_{2}\!\biggr)\!
+ \frac{47}{12}\gamma_{2}B_{1}
+ \frac{5}{2}\gamma_{3},
% \nonumber
\end{align}
\begin{align}
\Pi_{6,5} & =
\Pi_{1,0} + \frac{77}{60} \gamma_{1}B_{1} + \gamma_{2}.
% \nonumber
\end{align}

\begin{align}
\Pi_{7,4} & =
\Pi_{1,0}\biggl(\frac{85}{6}B_{1}^{2} + 10B_{2}\!\biggr)
+ \frac{57}{2}\Pi_{2,0}B_{1}
+ 15\Pi_{3,0}
+ \nonumber \\[1mm] &
+ \gamma_{1}\biggl(\frac{5}{8}B_{1}^{3} + \frac{23}{6}B_{1}B_{2} + 2B_{3}\!\biggr)
+ \gamma_{2}\biggl(\frac{59}{12}B_{1}^{2} + 5B_{2}\!\biggr)
+ \frac{37}{4}\gamma_{3}B_{1}
+ 5\gamma_{4},
% \nonumber
\end{align}
\begin{align}
% \nonumber
\Pi_{7,5} & =
\frac{87}{10}\Pi_{1,0}B_{1}
+ 6\Pi_{2,0}
+ \gamma_{1}\biggl(\frac{17}{6}B_{1}^{2} + 2B_{2}\!\biggr)
+ \frac{57}{10}\gamma_{2}B_{1}
+ 3\gamma_{3},
\end{align}
\begin{align}
% \nonumber
\Pi_{7,6} & = \Pi_{1,0} + \frac{29}{20}\gamma_{1}B_{1} + \gamma_{2}.
\end{align}

\begin{align}
\Pi_{8,4} & =
\Pi_{1,0}\biggl(\!\frac{105}{8}B_{1}^{3} + \frac{165}{4}B_{1}B_{2} + 15B_{3}\!\biggr)\!\!
+ \Pi_{2,0}\biggl(\!\frac{695}{12}B_{1}^{2} + 40B_{2}\!\biggr)\!\!
+ \frac{319}{4}\Pi_{3,0}B_{1}
+ 35\Pi_{4,0}
+ \nonumber \\[1mm] &
+ \gamma_{1}\biggl(\frac{19}{8}B_{1}^{2}B_{2} + \frac{59}{12}B_{1}B_{3}
+ \frac{29}{12}B_{2}^{2} + \frac{31}{12}B_{4}\!\biggr)
+ \gamma_{2}\biggl(2B_{1}^{3} + \frac{73}{6}B_{1}B_{2} + \frac{25}{4}B_{3}\!\biggr)
+ \nonumber \\[1mm] &
+ \gamma_{3}\biggl(\frac{89}{8}B_{1}^{2} + \frac{45}{4}B_{2}\!\biggr)
+ \frac{107}{6}\gamma_{4}B_{1}
+ \frac{35}{4}\gamma_{5},
% \nonumber
\end{align}
\begin{align}
\Pi_{8,5} & =
\Pi_{1,0}\biggl(\frac{413}{15}B_{1}^{2} + 15B_{2}\!\biggr)
+ \frac{459}{10}\Pi_{2,0}B_{1}
+ 21\Pi_{3,0}
+ \gamma_{1}\biggl(\frac{21}{8}B_{1}^{3} + \frac{33}{4}B_{1}B_{2} + 3B_{3}\!\biggr)
+ \nonumber \\[1mm] &
+ \gamma_{2}\biggl(\frac{139}{12}B_{1}^{2} + 8B_{2}\!\biggr)
+ \frac{319}{20}\gamma_{3}B_{1}
+ 7\gamma_{4},
% \nonumber
\end{align}
\begin{align}
% \nonumber
\Pi_{8,6} & =
\frac{223}{20}\Pi_{1,0}B_{1}
+ 7\Pi_{2,0}
+ \gamma_{1}\biggl(\frac{413}{90}B_{1}^{2} + \frac{5}{2}B_{2}\!\biggr)
+ \frac{153}{20}\gamma_{2}B_{1}
+ \frac{7}{2}\gamma_{3},
\end{align}
\begin{align}
% \nonumber
\Pi_{8,7} & = \Pi_{1,0} + \frac{223}{140}\gamma_{1}B_{1} + \gamma_{2}.
\end{align}

\end{document}